# FRAMEWORK FOR CLOUD COMPUTING ADOPTION: A ROADMAP FOR SMES TO CLOUD MIGRATION

Nabeel Khan and Adil Al-Yasiri

School of Computing Science and Engineering, University of Salford, Manchester, United Kingdom

## ABSTRACT

*Small and Medium size Enterprises (SME) are considered as a backbone of many developing and developed economies of the world; they are the driving force to any major economy across the globe. Through Cloud Computing firms outsource their entire information technology (IT) process while concentrating more on their core business. It allows businesses to cut down heavy cost incurred over IT infrastructure without losing focus on customer needs. However, Cloud industry to an extent has struggled to grow among SMEs due to the reluctance and concerns expressed by them. Throughout the course of this study several interviews were conducted and the literature was reviewed to understand how cloud providers offer services and what challenges SMEs are facing. The study identified issues like cloud knowledge, interoperability, security and contractual concerns to be hindering SMEs adoption of cloud services. From the interviews common practices followed by cloud vendors and what concerns SMEs have were identified as a basis for a cloud framework which will bridge gaps between cloud vendors and SMEs. A stepwise framework for cloud adoption is formulated which identifies and provides recommendation to four most predominant challenges which are hurting cloud industry and taking SMEs away from cloud computing, as well as guide SMEs aiding in successful cloud adoption. Moreover, this framework streamlines the cloud adoption process for SMEs by removing ambiguity in regards to fundamentals associated with their organisation and cloud adoption process.*

## 1. INTRODUCTION

The scalability and extensibility of distributed software architectures have led to the concept called Cloud Computing. Cloud computing is a technology used to deliver the hosted services over the Internet. Through this technology, users don't have to manage their own IT resources; instead they purchase their IT needs as services over the internet [1, 2].

For SMEs cloud computing technology is an attempt which has allowed them to reduce the rates spent on computing infrastructure. With more and more explorations of cloud technology, it has faced new challenges that need to be resolved to improve the pace of cloud adoption among SMEs.

With so much power of computation and data handling, there are numerous challenges that must be addressed promptly. Some of these challenges are technical (Interoperability and data security) and some are non-technical (Cloud knowledge and Cloud contracts); some of them are related to computer science whilst others are beyond computing realm. One of the issues which is hurting cloud adoption among SMEs are data security, privacy law and legal jurisdiction bound to certain boundary of the globe [3].

A major obstacle for SMEs in adopting cloud is how the software and hardware match and interact with each other while running the business applications. From a customer point of view it is an important factor to efficiently use the applications of cloud and mitigate any risks associated with it. This issue is often named as interoperability among clouds and addressing this issue is timely and necessary, and it's the only way to avoid vendor lock-in situation. So for SMEs to switch to cloud it is more important to know how to handle this issue as the market is dominated by big cloud providers. The knowledge of this potential issue is necessary as in case of multi provider scenario- SMEs can have plenty options to avail all the offered cloud services and at the same time takes the full advantage of cloud computing concepts like elasticity and pay per use model [4].

In order to start thinking of migrating organization to cloud adequate knowledge is a prime. SMEs usually have limited knowledge about what cloud can do for their organization in a long run and how it can enhance the productivity of its employees. Most SMEs IT professionals are not fully trained in guiding their organizations to cloud [5].

Although, cloud computing has been really successful in attracting customers, it is still considered as an emerging technology among the SMEs due to some challenges [6]. In order to find out these challenges and propose a guideline to SMEs, this paper aims to showcase the challenges related to cloud adoption by SMEs and propose a guideline to assist cloud transition from conventional IT rooms.

SMEs are not aware of cloud contract terms which are negotiated quite often by the cloud vendors. This issue is holding SMEs back to their conventional ways of doing business in the market, as they have fear of losing business by getting trapped into inconclusive terms of cloud contracts [7].

This paper explores possible solutions and elaborates on these issues. All these issues have emerged as a viable threat to cloud industry. Moreover, a stepwise guide in the form a framework has been proposed to aid cloud adoption process among SMEs.

The remainder of this paper is structured as follows. Section II describes the related work that has been done in this area along with contribution made by this paper. Section III describes the methodology and data analysis involved in making this research possible. Section IV states the key challenges found during interviews. Section V presents the details of proposed framework for cloud computing adoption. Finally, Section VI, VII and VII describe the further work to be conducted in this field, the conclusion and references respectively.

## 2. RELATED WORK WITH CONTRIBUTION

Several studies have been conducted on cloud issues and what is really affecting cloud industry. There are around more than 11 cloud frameworks in the market at present to deal with IT services or architecture but none of them is designed for cloud adoption for any size of an organization (this proposed cloud framework is only for SMEs in particular). Some of these frameworks are Cloud Business Model Framework (CBMF) by [8], IBM Framework for Cloud Adoption [9], Cloud Computing Business Framework [10] etc.

The cloud framework mentioned in this paper focuses on the readiness of the organization by making them aware of all the risks and preparedness associated with cloud computing and by proposing solutions to these. The contribution of this paper and eventually a cloud framework is in cloud computing adoption among small organizations, as it not only raises issues inflicting cloud computing but also provides remedy to avoid these issues to an extent.

## 3. METHODOLOGY AND DATA ANALYSIS

The research methodology in figure 1 has been adopted for this research program. This four-step guide (See Figure 5) is the vital part of the cloud preparation stage (CPS) of the proposed cloud framework. The CPS focuses onto the preparation part of the enterprises, as it will help them to understand the current situation of their business processes and guide them to progress in cloud adoption. The main stages of the methodology are outlined underneath:

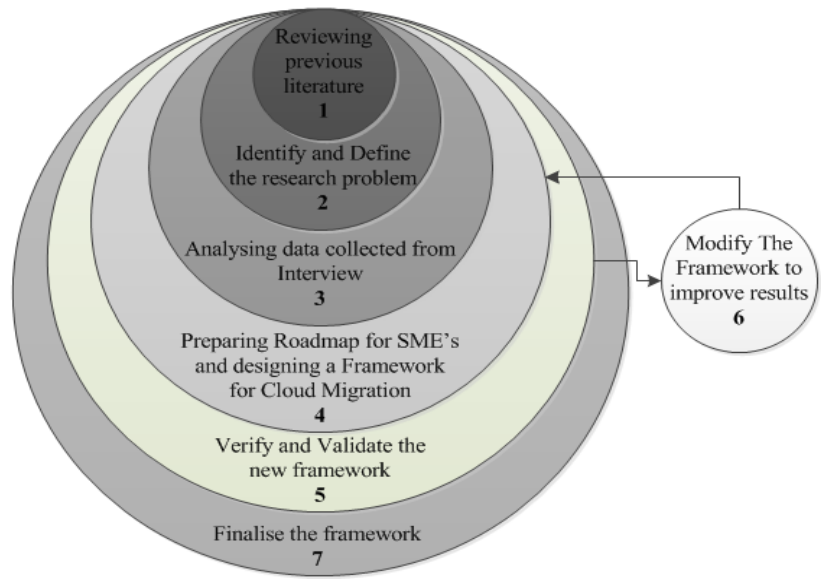

Figure 1. Research Methodology

1. Reviewing previous literature and relevant cloud computing issues: In this step, previous relevant literature, surveys and studies have been reviewed to identify issues with cloud adoption.
2. Identify and define the research problem: To do so, a qualitative research method has been adopted as a primary approach and following this; a number of semi-structured interviews were conducted with SMEs, cloud providers and developers to gather relevant information.
3. Analyzing data collected from interviews: The collected data from the interview is then analyzed by applying the content analysis approach.
4. Preparing a road map for SMEs: A structured guideline will then be designed in order to mitigate the cloud adoption barriers among SMEs.
5. The fifth step of the research methodology concentrates on verification and validation of the proposed framework and gathers a feedback from SMEs and Cloud Service Providers (CSPs). For this, reflexivity workshops will be held.
6. The feedback from the workshop will then be used to modify the framework to rectify and increase the efficiency of the framework.
7. Finally, the proposed framework will then be prepared for use by SMEs in cloud adoption.

The information required for the current research on finding cloud barriers among SMEs, has been collected by using primary research techniques and secondary research techniques. Interviews were the main sources of finding the primary information; where the source of secondary information gathering is based on the literature review method. A similar approach was adopted by [11] in 2012 where she interviewed several IT experts to find out cloud computing risks and solutions for rationalizing and mitigating these risks. Also, similar research was conducted by [12] in 2014 where he explored the factors influencing the adoption of cloud computing and the challenges faced by the business and it is also supported by [13] as this approach provides an in depth knowledge about the research problem.  Initially, in the first round of the interviews more than 60 SMEs were approached to know the issues that are keeping them away from adopting cloud computing. The results of these interviews are summarized as shown in figure 2. But during most of these interviews SMEs were perplexed over detailed knowledge of cloud computing. But some of the SMEs had an idea about the vulnerability of their data in cloud due to privacy and security issues with cloud computing. From the first round of the interviews, it was quite clear that the lack of expert knowledge about cloud computing is one of the main barrier for SMEs. Then a reversed approach was adopted and second round of interview surveys were conducted with eight cloud vendors and four other cloud developers to get the vital information on SMEs concerns over cloud adoption. And the sets of data collected after all these interviews were worth analysing as it unfolded some viable threats to cloud computing among SMEs.

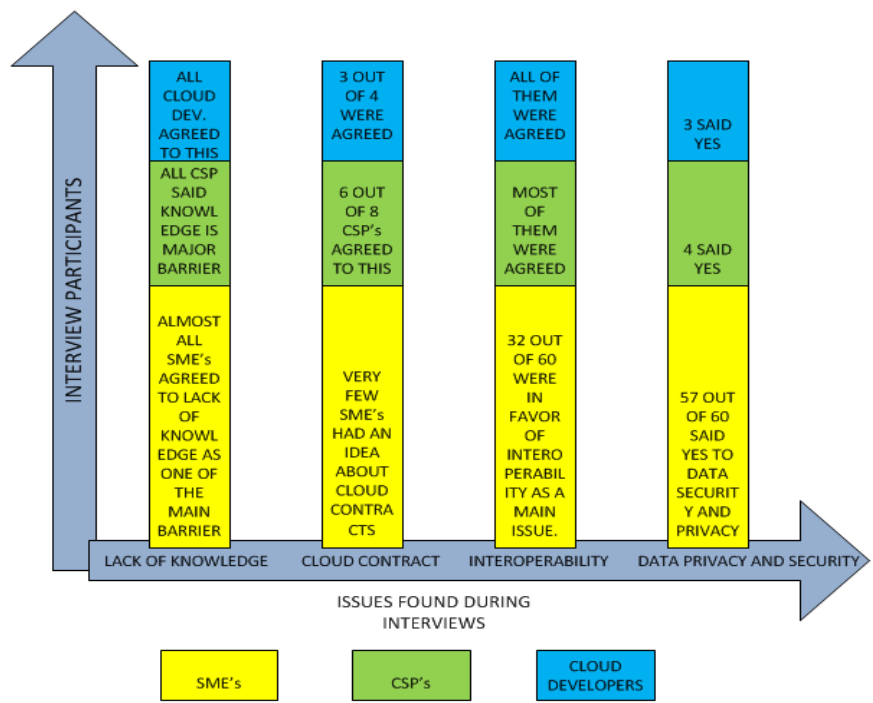

Figure 2. Interview results.

The content analysis is an approach applied in order to analyze the collected data (see figure 3). Firstly, the collected data was transcribed to avoid redundancy. It was then labelled according to the major attributes of cloud computing which makes the next step easy where data was divided into different themes. These themes were split into favorable attributes and non-favorable attributes of cloud computing in regards to SMEs. From the non-favorable attributes, several issues were identified which are hindering SMEs cloud adoption. These issues were then segregated into technical and non-technical issues in order to get clear understanding behind these constraints. Then a guideline to SMEs is proposed based on the findings from interviews, government standards, previous surveys and literature review.

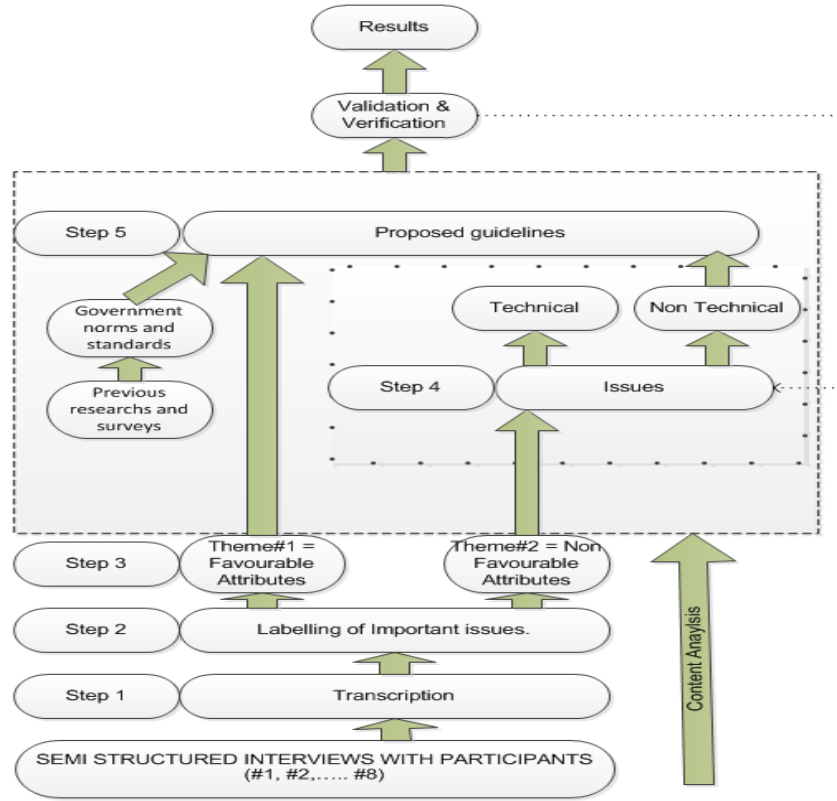

Figure 3. Data analysis.

For Validation and Verification purpose, reflexivity workshop will be conducted and all the previous participants (SMEs) will be invited to attend this workshop. During this workshop SMEs will be sharing their concerns with CSPs and will go through these four steps and framework for cloud adoption to validate how effective this framework is, in moving to clouds. Throughout this workshop comments from SMEs and CSPs will be noted to know how this framework is vital prior to any type of cloud adoption. During this workshop, the proposed framework will be validated to know its efficiency in bridging gaps between cloud computing and small organizations.

## 4. CHALLENGES FOUND DURING INTERVIEWS

Although, cloud computing has been successful in attracting customers, it is still considered as an emerging technology among the SMEs due to some challenges. There are a number of barriers hindering cloud computing adoption among SMEs. To identify and understand these, a pilot study was conducted in the form of interviews with the IT managers of eight leading cloud vendors and four cloud developers which have brought the following points into limelight:

- **Lack of internal staff expertise hindering cloud computing in transforming SMEs**: Slow uptake of cloud computing was attributed to the fact that SMEs need to be educated on the benefits and potential of cloud computing and correlating these benefits with their current IT requirements. When asked about the services which cloud vendors often offer to SMEs, all cloud vendors answered that they have to deal with an issue of SMEs being perplex over services or solutions that cloud computing can offer them. This was also inferred during the first round of interviews, as most SMEs were blank regarding cloud computing concepts (See figure 2). When asked about the in house cloud training to overcome this issue; SMEs said there are no such things in their company and cloud service providers (CSP's) answered that for most organizations there is no adequate training, no management and front end support for the cloud which according to CSP's are some serious issues to be addressed in cloud adoption [14, 15, 16].

- **SMEs concern over data and its safety**: The main concern for the SMEs is their data and its security. A question was posed on "how do cloud vendors deal with data security issues that SMEs have", CSP's replied, SMEs are over reluctant in giving full control of their data to cloud vendors. SMEs often complain about where their data is stored and what data jurisdictions protect it. When same question was asked to SMEs, 57 out 60, responded that Data security and privacy is one of the main issues (see figure 2). What happens in an advent of any security breach or snooping and how data will be recovered, are some of the issues that's hindering their cloud adoption.

- **SMEs prohibits classified data to be placed on shared or outsourced infrastructure**: This issue also plays an important role in inhibiting cloud migration by SMEs. When asked about time taken for data migration and its steps, most cloud developers and vendors replied that during migration interoperability is one of the main barrier for SMEs to cloud, as the problem arises when customers legacy applications, data and infrastructure; application programming interface (API) is not compatible with the API of the cloud services. Interoperability also emerges when a particular solution is provided on multiple cloud models and platforms like a solution provided partially on IaaS service mode and remaining on SaaS or on different deployment layers like public and private. Interoperability also appears when different CSP's are involved in giving solution to an organization [4]

  When the question about interoperability was asked to SMEs, quite a few of them were able to answer this question as they had little or no idea regarding this issue. But when explained about interoperability and how does it occur, more than half of them agreed to it, as shown in figure 2.

- **SMEs are perplexed over regulatory and service terms in cloud contracts**: The fourth most predominant problem found during the interviews is the cloud contract. When asked about what SLA is on offer and how do cloud vendors deal with incompliance issues in cloud contract, cloud vendors rated this as among the most viable challenges to cloud computing. On the other hand, SMEs were perplexed due to complicated terms and conditions of cloud contracts and lack of knowledge regarding cloud concepts. Also, cloud vendors said we dictate terms while signing contracts as SMEs have little idea about vital terms and conditions related to cloud contracting. The most common cited terms found are liability, SLA's, security, liability, and privacy. [17].

The unawareness of these issues has blotted cloud computing among SMEs which needs to be cleansed by taking some necessary steps.

## 5. FRAMEWORK FOR CLOUD COMPUTING ADOPTION

The challenges identified above affect cloud adoption among SMEs which need to be addressed with adequate solutions and recommendations. All the above problems are directly related to a bigger problem; which is the lack of framework for SMEs that directs the process of migration to clouds. Eventually, a framework for SMEs could offer an answer to such problems in a structured manner to expedite the cloud adoption among SMEs. This section provides the design and explanation of the components of cloud framework. The framework in this research is the stepwise guide for the SMEs to follow their journey on cloud adoption. This framework is divided in three stage migration process; cloud preparation stage, cloud requirement stage and cloud migration stage as shown below in figure 4.

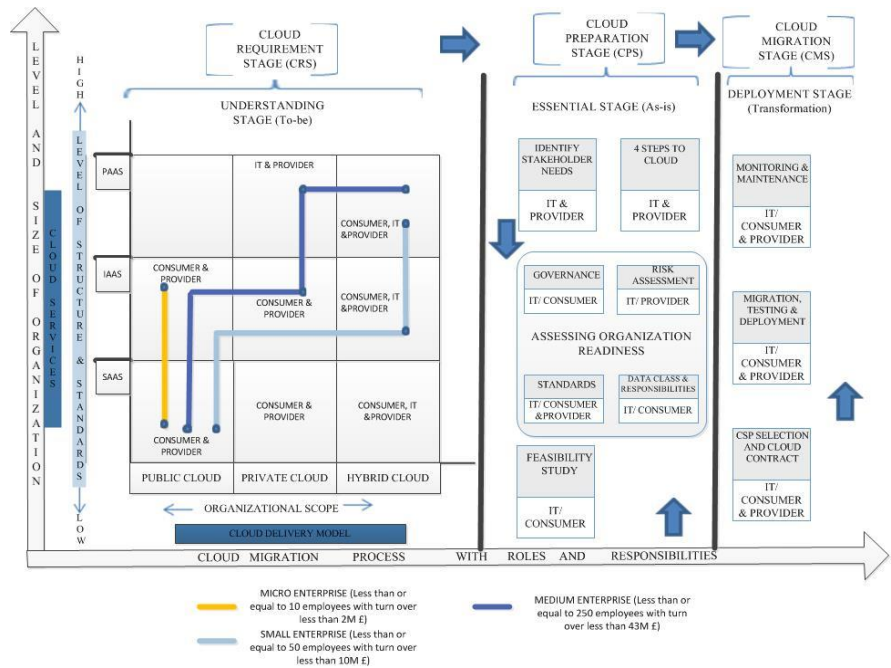

Figure 4. Framework for Cloud Computing Adoption.

The cloud computing process of migration requires a continuous process of improvement prior to migration to ensure the efficiency of the system (or solution) and for this framework works in a circular loop to provide firm and positive results. For this As-Is and To-Be business process is adopted where Cloud Requirement stage (CRS) of the framework is To-be, Cloud Preparation Stage (CPS) is As-is and Cloud Migration Stage (CMS) is transformation. As the CPS goes further, the goal should be set to prepare the enterprise for cloud computing adoption and regular modification needs to be made to refine the process of getting SME ready for cloud computing. These refinements will be vital in deciding the efficiency and sustainability of the system. As the process enters in a CMS stage, it is ready to be transformed as per the proposed solution provided (or selected) to an enterprise.

### 5.1 CLOUD REQUIREMENT STAGE (CRS)

This stage elaborates regarding cloud computing services and deployment models that may be required by SMEs depending upon their head count, turn over and nature of their businesses. All the depictions in this stage are based upon market study and cloud vendor's advice for SMEs. Based on the understanding, consumers or users will be able to assess the solution that they would need in order to meet their requirements.

SMEs with less than or equal to 10 employees and turnover of less than 2 million pounds (called micro enterprise), would only go for either software as a service or Infrastructure as a service over a public cloud model. Whereas SMEs with less than or equal to 50 employees and with turnover of less than 10 million pounds, usually embraces IaaS as a standalone service deployed over public cloud environment and in some cases they go for amalgamated services over hybrid cloud. SMEs with head count less than or equal to 250 and turnover of less than 43 Million pounds, either go for only IaaS or PaaS on a private cloud or it may go for collection of services (SaaS, IaaS, PaaS or mix) on mixed cloud environments.

### 5.2 CLOUD PREPARATION STAGE (CPS)

This is the second stage of the cloud migration process and it is called the cloud preparation stage (CPS). It focuses onto the preparation part of the enterprises, as it will help them to understand the current situation of their business processes and guide them to progress in cloud adoption.

#### 5.2.1 Feasibility Study

This usually comprises of the process of finding whether migration to cloud computing is suitable for an organisation, in both terms; financial and technical terms. This very first step in the transition to cloud involves the study of the existing data, the analysis of existing applications running in the organisation, the requirement analysis of the data for a desired output from the system, and study on any constraints that may be faced pre and post cloud migration. It also contains the detailed examination of the hardware available within the organisation in order to perceive the additional costs incurred in terms of hardware for embracing cloud. It's highly recommended to use feasibility tools, like a maturity assessment tool, technical feasibility tool or business feasibility tool [18]. These tools give an insight about the technical and business feasibility of the organisation and helps in evaluating the effort and cost required in the process of migration to the cloud.

#### 5.2.2 Assessing Organisation Readiness

It's important to know where an organisation stands in terms of preparedness to embrace the cloud technology. The followings are four vital factors which covers the readiness of an organisation: Governance, Risk Assessment, Standards and Data classification and responsibilities.

##### 5.2.2.1 Governance

This part of the organisational assessment is all about making wise (business oriented) decisions when it comes to predicting performance and taking responsibilities among the organisation. It's basically, how well the policies and rules should be implemented to ensure the smoothness in the operation of an organisation. Prior to migration organisations should ensure IT governance is implemented. IT governance helps in enhancing performance, confidentiality, risk management, strategies and resources. IT governance is made from members of business and IT division within the company. Usually IT governance takes care of IT assets, ensuring approved running of all hardware systems, process etc. as per the policies and procedures of the company. IT governance looks after the maintenance and control of available assets, also, ensuring these assets contribute in the organisations business strategy and goals. [19]

##### 5.2.2.2 Risk Assessment

During interviews several technical and organisational risks were identified and they are categorised depending upon their effects on the organisation. Also, these risks have been formulated from the previous researches made on cloud computing risks.

Table 1. Representation of risk assessment table.

| 1.Probability | 2.Impact | 3.Vulnerability | 4.Affected assets | 5.Risk inference |
| --- | --- | --- | --- | --- |

1. The level of probability and effects of these risks are calculated based upon the expert reviews of cloud developers and previous researches (see table 1). Levels used in risk assessment are Often, Rare and Not Applicable (N/A).
2. The level of impact on small business has been determined in assistance of cloud experts (see table 1). Levels used in risk assessment are Severe, Medium and Low.
3. Vulnerabilities have been identified using interviews with cloud developers and it basically describes about the reason which may lead to these risks (see table 1).
4. Affected assets have been mostly found from the previous researches and cloud developers guidance has also been taken. This column shows effect of risks on assets (see table 1).
5. Risk inference is the overall conclusion of the research made on risk with level of risks as high, medium and low respectively. The level of underlying risks has been identified on the basis of incident occurrences in the past (see table 1).

Depending on area of impact the possible risks have been categorized as follows:

### 5.2.2.2.1 Policy And Organizational Risks

Risk# 1 Loss of governance: The loss of governance may impact the organization in terms of security, compliance, data availability, integrity, confidentiality, poor performance and service. According to study conducted by [20], this risk can be minimized mainly through adopting certain measure within the organization such as: Clear roles and responsibilities, appropriate SLA clauses, Adequate use of technology and solutions provided and avoiding ambiguity over terms of use.

Risk#2Compliance: Moving to cloud computing needs certain type of compliances and these compliances can be used to either advertise it as a brand feature, meet regulatory requirement or industry standard. The investment incurred in getting these compliances is hefty and can be at risk if cloud service provider (CSP) has no such compliance evidence according to their requirements. SMEs must ensure the availability of these compliances prior to moving to cloud.

### 5.2.2.2.2 TECHNICAL RISKS

Risk#3 Insider Abuse of Privilege risk: This risk may disrupt all kinds of services, may have an adverse on data confidentiality, availability and integrity, and may blot overall organization's reputation and customer's trust. SMEs must ensure the availability of intrusion detection and incident response controller at the cloud service provider's end with clear roles and responsibilities. [20]

Risk#4 Data leakage while uploading or downloading data risk: In Cloud computing, data is always on the move either within or between the clouds or between available infrastructures at clients place. Data in the cloud is always transferred among multiple distributed images and these images are distributed across multiple machines of customers', between cloud infrastructures and remote web clients etc. SMEs must ensure that this hosting from data centers is done using a secure Virtual Private Network (VPN) connection.

Risk#5 Distributed denial of service (DDoS), Economic denial of service (EDoS) and data deletion risks: DDoS occurs where an attacker attacks customer's metered resources using public channel. EDoS when cloud resources are used to disable economic drivers of using cloud infrastructure services. In identity theft, an attacker may use customer data, account or resources for personal gains or to spoil customer's image. SME's must ensure that CSP is using Distributed cloud intrusion detection model, confident based filtering, virtual machine monitoring, EDoS-shield etc. mechanisms to prevent these attacks from happening while offering services.

### 5.2.2.2.3 LEGAL RISKS

Risk#6 Court Summons and E-discovery risk: Client's data is at risk in the event of hardware confiscation by law agencies or civil bodies, in this case other non-associated identities are at risk of being revealed to unwanted people. SMEs must ensure to avoid this risk with an appropriate clause in the cloud contract as this results in disclosure of other client's data too.

Risk#7 Change of Jurisdiction and data protection Risk: Multiple locations (or jurisdictions) may be used to store customer's data, if location of the data center is in a country where law and order framework is unstable and with no adequate data protection laws to protect the customer's data, where international agreements are mistreated and sites where data centers can be raided enforcing data disclosure. In these cases SMEs must ensure CSPs data centers are in UK & Europe or countries in which there are separate cloud computing laws already in practice. On the basis of jurisdiction, any incompliance to data protection laws may result in civil, administrative and criminal sanctions for both, data controller and cloud provider, as their duties are different when it comes to data processing and handling. The best practice is to always seek for cloud providers who provide certifications on their data processing and data security activities like SAS70 (Statement on Auditing Standards No. 70) certification provider.

Risk#8 Risk associated with licensing in cloud: Licensing is also one of the big risk which affects overall structure in cloud computing, per seat agreements, number of instances per use and online licensing check are some of the licensing conditions which may become ineffective in cloud environment. SME's must ensure that CSP offers adequate licensing schemes pertaining to number of users.

### 5.2.2.3 STANDARDS

In a process of embracing cloud technology SME should either ask a provider for standards or implement certain standards within. Some of these standards are not obligatory to be implemented by every SME, as it varies from size and nature of the business that they are in. However it's good to have best practices in place while going for a new technology. These standards have been divided into security and confidentiality.

### 5.2.2.3.1 SECURITY

1. ISO/IEC27001: This standard enables organizations to secure their information assets.
2. CSA's Cloud Controls Matrix (CCM) and Consensus assessments initiative questionnaire (CAIQ): CAIQ helps in overcoming the leading concern about lack of transparency regarding data protection and risk management practices of cloud providers while implementing a solution. CCM is a step forward or an implementation of a CAIQ, as this contains the requirements and controls that a company wants from their cloud provider. In this a customer can easily go back and forth to prepare a set of requirement and control guidelines for the cloud vendor.

### 5.2.2.3.2 CONFIDENTIALITY

1.Transport Layer Security (TLS): Internet Engineering Task Force (IETF) is an open group for all users, developers and engineers which ensures the smooth running of the internet and they created protocol TLS. TLS is a protocol which ensures the communications done on computer network are secure and safe.

2.OASIS Key Management Interoperability Protocol (KMIP): It's basically a protocol through which cryptographic keys are manipulated by defining a certain message format, on a key management server. Cryptography is used to retain the confidentiality of a data, by coding and decoding the messages using keys which are known to sender and recipient of the message.
3.Media Sanitization (SP- 800-88): This standardization talks about how different sorts of media is sanitized before disposal and reuse. [21]

### 5.2.2.4 DATA CLASSIFICATION AND RESPONSIBILITIES

Data classification is all about segregating data and tagging it, which enables enterprises and cloud providers find the data quickly and efficiently. The types of terminologies used in the classification of data are; confidential, secret, top secret, for internal use only and private depending upon the nature of the data. The owner of the data has to be clear about the sensitivity of the data and its content. Before the data migration, it is the responsibility of an organization to classify their data, and the cloud provider should give a written commitment to an enterprise on how they maintain the privacy and security of their data in their cloud.

### 5.2.3 IDENTIFY STAKE HOLDER NEEDS

There are two major stakeholders in cloud computing, which are providers and consumers (or Users). In stakeholder perspective, the cloud providers are a group of stakeholders who perform maintenance, deliver services and upgrade the system to ensure the trouble free and secure running of the solution provided.

From the enterprise perspective, stakeholder needs are mostly comprised of their user's needs which encompasses in business productivity. The decision makers of a cloud program should use especially designed stakeholder needs platforms and questionnaires to identify users need.

### 5.2.4 FOUR STEPS TO CLOUD

There are number of barriers hindering cloud computing adoption among SMEs. To identify and understand these, a pilot study was conducted in the form of interviews which are mentioned below in figure 5 (Also, discussed in section 4).

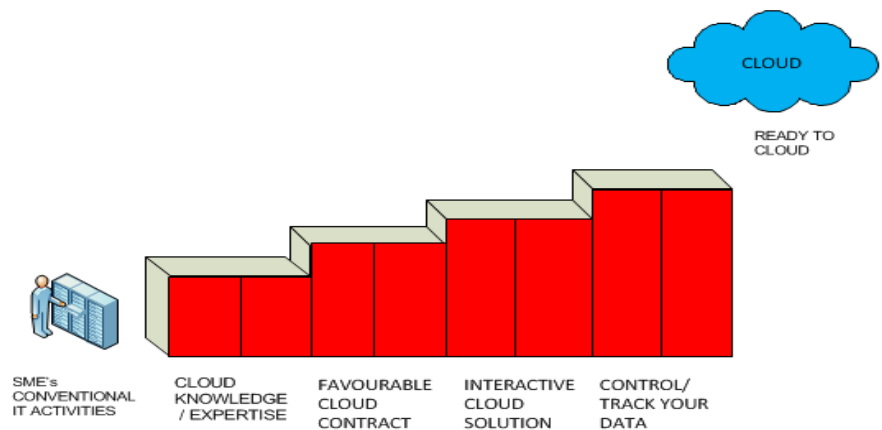

Figure 5. Four steps to get ready for cloud.

The challenges identified above affect cloud adoption among SMEs which needs to be addressed with adequate solutions and recommendations which are as follows:

1. <u>Gain cloud knowledge and seminate it within the organization for successful cloud execution:</u> SMEs should invest to have dedicated budget for cloud training courses for the employees. A set pre requisite for prospective employees to have cloud knowledge and formal request to educational institutes to introduce cloud computing courses. SMEs should market their success stories and contract pre & post cloud migration to encourage cloud adoption among other SMEs.[22]
2. <u>Know what you want and dictate your cloud contract:</u> SMEs should be well equipped in cloud computing concepts and seek a cloud provider who is more open and transparent towards appropriate standards and certifications which could be accepted by legislators and regulators. SMEs should also let insurers get involved while signing a contract, as insurers may be better able to assess the risks. [17]
3. <u>Care before you share, to avoid interoperability issues among the systems</u>: One of the most formidable and reliable solutions to resist interoperability issue within cloud or on premises is MuleSoft. SMEs must ensure that CSP has; Open Cloud Computing Interface, Webs Service Agreement Specification, Cloud Infrastructure Management Interface, Open Virtualization Format and Cloud Data Management Interface; standardizations prior to cloud migration to narrow the extent of interoperability issues among cloud services and legacy systems. [23, 24, 25, 26]
4. <u>Know who sees your data and where it's stored:</u> SMEs must ensure that any data migrated to cloud should have an approach which must incorporates encryption, key management, strong access control and security intelligence for the sake of protection of data and providing sufficient level of security. SMEs must know where there data is being hosted at all times as data in cloud is mobile at all times and they should also ensure that data location is in UK or EU. Also, they can ask for devices to track the location of the data at all times. SMEs may also ask cloud vendors to host their data on elastic data warehouses which provide high performance, elasticity and multi tenancy. [27]

## 5.3 CLOUD MIGRATION STAGE (CMS)

The stage describes about the migration of SME to cloud computing and going live process.

### 5.3.1 CLOUD SERVICE PROVIDER AND CLOUD CONTRACT:

This section of CMS phase basically portrays attributes to look for in a Cloud service provider after internal preparation is done by SME. These attributes are: assure that CSP is transparent in pricing when it comes to subscriptions, pay as you go models, upgrades, maintenance, exit cost and any other liabilities, look for CSP who is good in scalability and flexibility in clouds and prefer CSP who has all the good security practices in place in order to provide services to their customers. For cloud contract please refer to four steps to cloud.

### 5.3.2 MIGRATION TESTING AND DEPLOYMENT:

After the study and evaluation all the agreed applications or databases, it will now be migrated slowly and then tested. Prior to going live, it is essential that system is tested on regular bases with production data already migrated. Ensure parallel operation approach is adopted in order to avoid any unforeseen events which may lead to system error.

#### 5.3.2 MONITORING AND MAINTENANCE:

The importance of this stage can be realised as most issues erupts only after, when system goes live and CSP is in control of everything. The data or application over the cloud needs to be monitored for its availability, performance and security. There are various monitoring tools provided by the CSP, as these tools also helps in monitoring the compliance of SLA's and maintain the contract as per the agreement signed between customer and CSP.

## 6. FURTHER WORK

The next stage of the research is to validate and verify the proposed framework and for this purpose all the previous participants will again be invited to attend the reflexivity workshop where they will be explained about this framework to know the validity and effectiveness of this framework. SMEs and CSPs will be in a discussion regarding how helpful this framework can be in aiding cloud adoption among SMEs. Moreover, the purpose of this workshop is to know about the pre & post migration issues that SMEs face when they approach to CSPs, the collected data here will be useful in modifying a cloud framework for SMEs.

## 7. CONCLUSION

The usage of cloud computing is slow among SMEs, as SMEs require services more in the area of offering infrastructure and software as a service. A number of issues were discussed above, which have a great impact on cloud adoption among small and medium size enterprises. The study and analysis of these issues have led to proposition of framework for cloud computing adoption, which give the understanding and a roadmap to SMEs on how to approach cloud vendors or specifically how to adopt cloud computing and minimize potential barriers. Such a framework will ease up cloud migration by assessing organization's readiness for the cloud through parameters like governance, risks involved, standards, data classification and responsibilities. Also, it will enable organizations to identify stakeholder needs, select right CSP and sign a favourable cloud contract, which will help SMEs in a smooth transition to cloud.

We believe that this framework will encourage and accelerate the adoption of cloud computing among such enterprises.

## 8. FUTURE RESEARCH

We believe that research in the field of framework for cloud computing migration would benefit substantially if this framework is implemented in a real time environment where enterprises use this framework as an aid to their cloud migration process and provide feedback with case study. This could provide practical examples on framework implementation in real time and its effectiveness.

International Journal on Cloud Computing: Services and Architecture (IJCCSA) ,Vol. 5,No. 5/6, December 2015

**AUTHORS**

I (**Nabeel Khan**) graduated from the Birla Institute of technologies and Sciences, Pilani in 2010 with a BSc in Computer Science, and completed my MSc in Business Computing from Birmingham City University in 2012. Currently I am pursuing my PhD in Cloud Computing from University of Salford, Manchester and expected completion is by 2016. 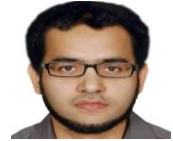

I (**Dr Adil Al-Yasiri**) graduated from the University of Technology (Baghdad) in 1984 with a BSc in Systems and Control Engineering, and completed my MSc in Instrumentation and Control Engineering from the same university in 1988. I worked as a Instrumentation engineer before completing a PhD on Domain Oriented Object Reuse using Generic Software Architectures in 1997 from Liverpool John Moores University (UK). After 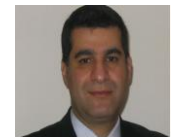 completing my PhD I lectured at universities in the UK and the Middle East, and became a head of Computer Science at UAE University (Al-Ain).Between 2000 and 2003, I worked on number of projects around the world advising clients (nationally and internationally) on various aspects of the software development process. In December, 2003 I joined the University of Salford as a lecturer in computer network systems and became senior lecturer in 2009. I am currently the programme leader for the MSc software engineering course.